\pdfoutput=1
\documentclass[10pt, conference]{IEEEtran}
\usepackage{cite}

\ifCLASSINFOpdf
  \usepackage[pdftex]{graphicx}
\else
  \usepackage[dvips]{graphicx}
\fi
\usepackage{url}

\usepackage{listings}
\usepackage{color}
\usepackage{subcaption}
\usepackage{booktabs}
\usepackage{amssymb}
\usepackage{amsmath}

\newcommand*\rot{\rotatebox{90}}

\definecolor{mygreen}{rgb}{0,0.6,0}
\definecolor{mygray}{rgb}{0.5,0.5,0.5}
\definecolor{mymauve}{rgb}{0.58,0,0.82}
\definecolor{light-gray}{gray}{0.95}

\usepackage{xargs}
\newcommandx{\unsure}[2][1=]{}
\newcommandx{\change}[2][1=]{}
\newcommandx{\info}[2][1=]{}
\newcommandx{\improvement}[2][1=]{}
\newcommandx{\thiswillnotshow}[2][1=]{}

\usepackage{draftwatermark}
\SetWatermarkText{Pre-print}
\SetWatermarkScale{1}
\SetWatermarkLightness{0.85}

\IEEEoverridecommandlockouts
\IEEEpubid{\makebox[\columnwidth]{ 978-1-5090-0933-6/16/\$31.00~\copyright~2016 IEEE \hfill} \hspace{\columnsep}\makebox[\columnwidth]{ }}

\begin{document}
%
\title{MeDICINE: Rapid Prototyping of Production-Ready Network Services in Multi-PoP Environments}

\author{\IEEEauthorblockN{Manuel Peuster}
\IEEEauthorblockA{Paderborn University\\
manuel.peuster@uni-paderborn.de}
\and
\IEEEauthorblockN{Holger Karl}
\IEEEauthorblockA{Paderborn University\\
holger.karl@uni-paderborn.de}
\and
\IEEEauthorblockN{Steven van Rossem}
\IEEEauthorblockA{Ghent University – iMinds, INTEC\\
steven.vanrossem@intec.ugent.be}
}


%


\maketitle

\begin{abstract}

Virtualized network services consisting of multiple individual network functions are already today deployed across multiple sites, so called multi-PoP (points of presence) environments. This allows to improve service performance by optimizing its placement in the network. But prototyping and testing of these complex distributed software systems becomes extremely challenging. The reason is that not only the network service as such has to be tested but also its integration with management and orchestration systems. Existing solutions, like simulators, basic network emulators, or local cloud testbeds, do not support all aspects of these tasks.

To this end, we introduce \emph{MeDICINE}, a novel NFV prototyping platform that is able to execute production-ready network functions, provided as software containers, in an emulated multi-PoP environment. These network functions can be controlled by any third-party management and orchestration system that connects to our platform through standard interfaces. Based on this, a developer can use our platform to prototype and test complex network services in a realistic environment running on his laptop.

\end{abstract}



%
\IEEEpeerreviewmaketitle

\section{Introduction \& Motivation}
\label{sec:intro}


The emerging trend of network function virtualization (NFV) promises a new level of flexibility for the upcoming 5th generation of networks. It turns network functionality, previously implemented as proprietary hardware boxes, into software artifacts that are executed in virtualized environments. Multiple of these virtual network functions (VNF) are then connected and create complex network services (NS), which are controlled by a management and orchestration (MANO) system \cite{karl2016devops} \cite{unify2015}. Such orchestrators do not only manage the deployment of network services but also automate operational management, e.g., service scaling.

Virtualized network services can be distributed in the network and different functions of a service can be executed in different points of presence (PoPs). These PoPs may be full-fledged data centers but also smaller sites, like network edge routers or base stations, that offer a limited amount of compute resources to run arbitrary functions.

In such an environment, the creation of network services becomes a complex software development process that consists of two main parts. First, the development of the service and its functions as such. Second, the integration of the service with a MANO system that manages the service during its lifecycle. The second part involves the implementation and test of management interfaces but also the design, implementation, and validation of service-specific management components, like auto-scaling rules or placement strategies~\cite{karl2016devops}. This complicates the overall development process. To reduce this complexity, extended tool support is required to reduce time-to-market, save costs, and improve the quality of the developed services.

A special problem in this process is the lack of supporting tools to locally prototype or test complete network services in end-to-end multi-PoP scenarios. These tools should allow testing a  network service as such, e.g., by sending generated traffic through it, but also validating its interaction with a MANO system, e.g., dynamic reconfiguration or placement strategies. This is not possible with existing approaches which either rely on local cloud testbeds that lack multi-PoP support, simulations that only execute simplified versions of network functions, or network emulation tools that do not offer interfaces to interact with MANO systems.

There are both simple and complex use cases for such a development support tool and we present some of them to motivate our proposed solution. These use cases can be divided into two categories. First, use cases that check the functionality of the network service as such (\emph{NF-UC}); this can already be done with existing emulation solutions but requires considerable manual effort, e.g, for adapting the emulated services for production environments. Second, use cases that check the interoperation between network service and a MANO system (\emph{MANO-UC}), which can today only be done with complex cloud testbeds. We list here some examples for both categories.

\paragraph{\textbf{NF-UC1 (Single VNF)}} A network service developer wants to deploy and test single VNFs in a local test environment by sending some generated traffic through them. Such VNFs should be executed as containers so that the same container images can directly be deployed in a production environment. During the test, a developer wants to interact with the running VNFs to, e.g., change configurations or monitor their behavior.

\paragraph{\textbf{NF-UC2 (Complex services)}} A developer wants to test entire complex services consisting of several chained VNFs. 
A local test environment should be able to execute such complex services so that end-to-end tests, e.g., sending traffic through the service's chain, can be performed.

\paragraph{\textbf{MANO-UC1 (Service management)}} There is the need to validate the service behavior in dynamic deployments in which the service is modified at runtime by a MANO system. To do so, a test environment needs to be able to interface with existing MANO systems to manage the executed services just like in production environments. An example is to test reconfiguration mechanisms after adding additional VNF instances to the service (scale-out). Such tests should be performed in multi-PoP environments to also allow tests of placement optimization strategies and service management across multiple PoPs.


\paragraph{\textbf{MANO-UC2 (Feedback-based autoscaling)}} A network service developer wants to test custom autoscaling approaches. To support this, connected MANO systems have to be able to collect feedback data, e.g., monitoring information, from services executed in the test platform as well as from the platform as such.

\begin{figure}[t]
	\centering
	\includegraphics[width=0.7\columnwidth]{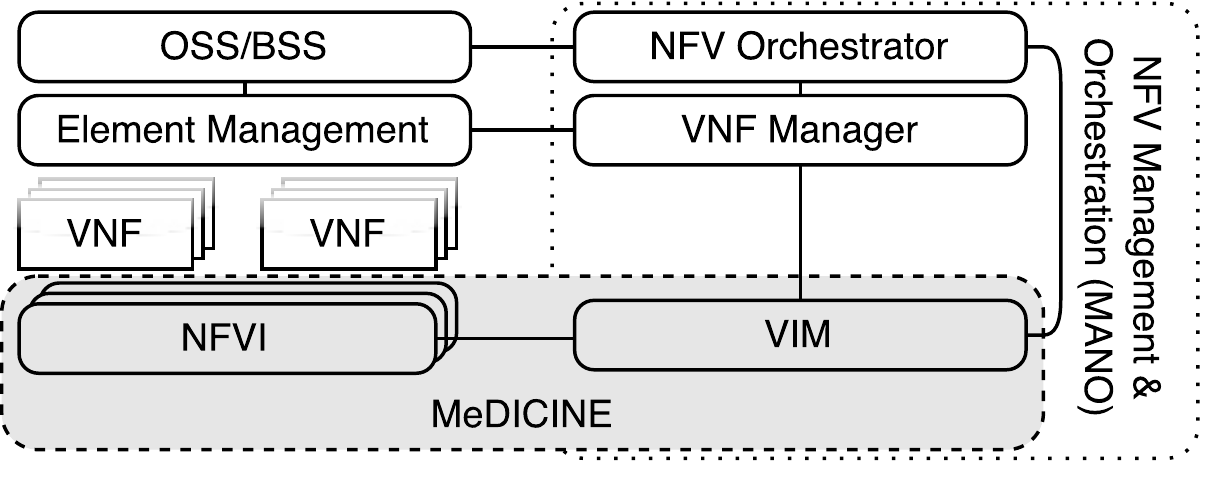}
	\caption{The MeDICINIE platform in the simplified ETSI NFV reference architecture \cite{etsi2network}.}
	\label{fig:etsi_mapping}
\end{figure}

To cover the previously described use cases and overcome the shortcomings of existing development support tools (Sec.~\ref{sec:rw}), we introduce \emph{MeDICINE} (Multi Datacenter servIce ChaIN Emulator), a novel prototyping platform for network services. Our platform is able to execute production-ready network functions in realistic multi-PoP environments and allows standard MANO systems to control the deployment, like in a real-world system. Fig.~\ref{fig:etsi_mapping} shows the scope of our solution and its mapping to the ETSI NFV reference architecture in which it emulates the network function virtualization infrastructure (NFVI) and the virtualized infrastructure manager (VIM). 

The remainder of this paper is organized as follows. First, we compare existing simulation, emulation, and testbed solutions in Section~\ref{sec:rw} and describe our platform in Section~\ref{sec:medicine}. Section~\ref{sec:eval} presents first experimental results and demonstrates the platform. Section~\ref{sec:conclusion} concludes.

\section{Related Work}
\label{sec:rw}

NFV development support is still a novel research direction with a limited amount of existing solutions, most of them focusing on SDN debugging rather than on prototyping of complex network services \cite{pelle2015}. Other approaches are based on simulations to test and validate management solutions, e.g., placement algorithms, but they only provide very limited realism since the simulated network functions are only proxies and not real implementations used in production \cite{calheiros2011cloudsim} \cite{simulationoverview2012} \cite{henderson2008network}. VLSP \cite{mamatas2015service} offers more realism but the tested network functions are still limited to be simple Java programs and not real NF implementations.

Emulation tools, like Mininet, are able to execute any network function implementation in its own virtualized network namespace \cite{lantz2010network} \cite{maxinet2014} \cite{core2010}. However, moving these network functions into a production environments is still a time-consuming task and these tools lack the possibility to emulate PoPs or cloud sites, e.g., they have no functionality to stop and remove hosts at runtime. The ESCAPE platform overcomes some of these limitations by combining a MANO system with multi-PoP support (including Mininet and OpenStack) but it does not target development support or prototyping tasks. Its main focus is on orchestration between non-emulated PoPs~\cite{unify2015}.

Real cloud testbeds, which might be installed on a single physical machine, are typically not able to run services in arbitrary network topologies \cite{OpenStackDevStack.wepage}. And even if they do, they come with considerable management overhead and only a limited number of PoPs that can be used for tests \cite{keller2013}.

%

\begin{table}[h]
\centering
\caption{Feature matrix of existing approaches}\label{table:compare}
    \begin{tabular}{@{} cl*{7}c @{}}
        & & \rot{Simulations} & \rot{VLSP \cite{mamatas2015service}} & \rot{Mininet \cite{lantz2010network}} & \rot{ESCAPE \cite{unify2015}} & \rot{DevStack \cite{OpenStackDevStack.wepage}} & \rot{Cloud testbeds}  & \rot{MeDICINE} \\
        \cmidrule{2-9}
        & production ready NFs	& - & - &  o &  + &  + & + & + \\
        & multi PoP  					& + & + &  o &  o &  - & o & + \\
        & arbitrary topologies 			& + & + &  + &  o &  - & - & + \\
        & realistic NF performance  	& - & - &  o &  + &  o & + & o \\
 		& explicit chaining support		& o & o &  - &  + &  - & o & + \\
	 		               \rot{\rlap{~Feature}}
        & MANO system integration  		& o & o &  - &  + &  + & + & + \\
        & run offline/local			  	& + & + &  + &  o &  o & - & + \\
        & test/prototyping support		& + & + &  + &  - &  o & o & + \\
        \cmidrule{2-9}
        \multicolumn{9}{c}{+ : fully supported, o : partly supported, - : not supported}\\
        \cmidrule{2-9}
    \end{tabular}
\end{table}

Table~\ref{table:compare} presents an overview of features needed for NFV development support and shows which of them are provided by existing simulation approaches, emulation tools, and testbed solutions. It shows that all existing approaches lack some features and that \emph{MeDICINE} is the only solution that can execute production ready NFs in a emulated multi-PoP environment on small amounts of hardware, like a developer's laptop.

\section{MeDICINE Platform}
\label{sec:medicine}

\begin{figure*}[t]
	\centering
    \begin{subfigure}{.60\linewidth}
  		\centering
  		\includegraphics[width=0.9\linewidth]{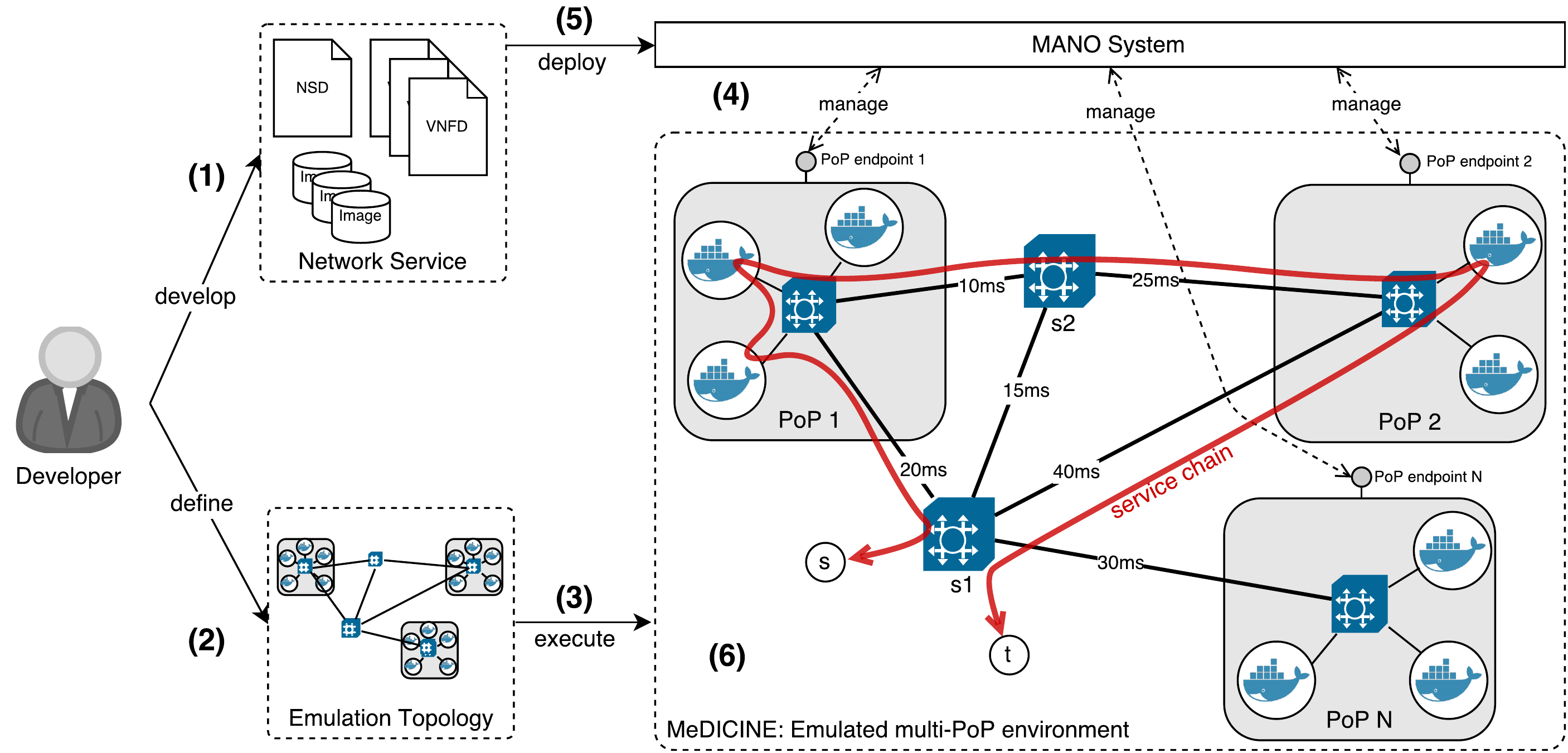}
  		\caption{General idea and workflow of the system. Example of a running emulation environment with three PoPs, eight allocated compute instances executing VNFs, and a service chain setup consisting of three chained VNFs distributed across two PoPs through which generated traffic is sent from $s$ to $t$.}
  		\label{fig:medicine-idea}
	\end{subfigure}\hfill
	\begin{subfigure}{.30\linewidth}
  		\centering
  		\includegraphics[width=0.9\linewidth]{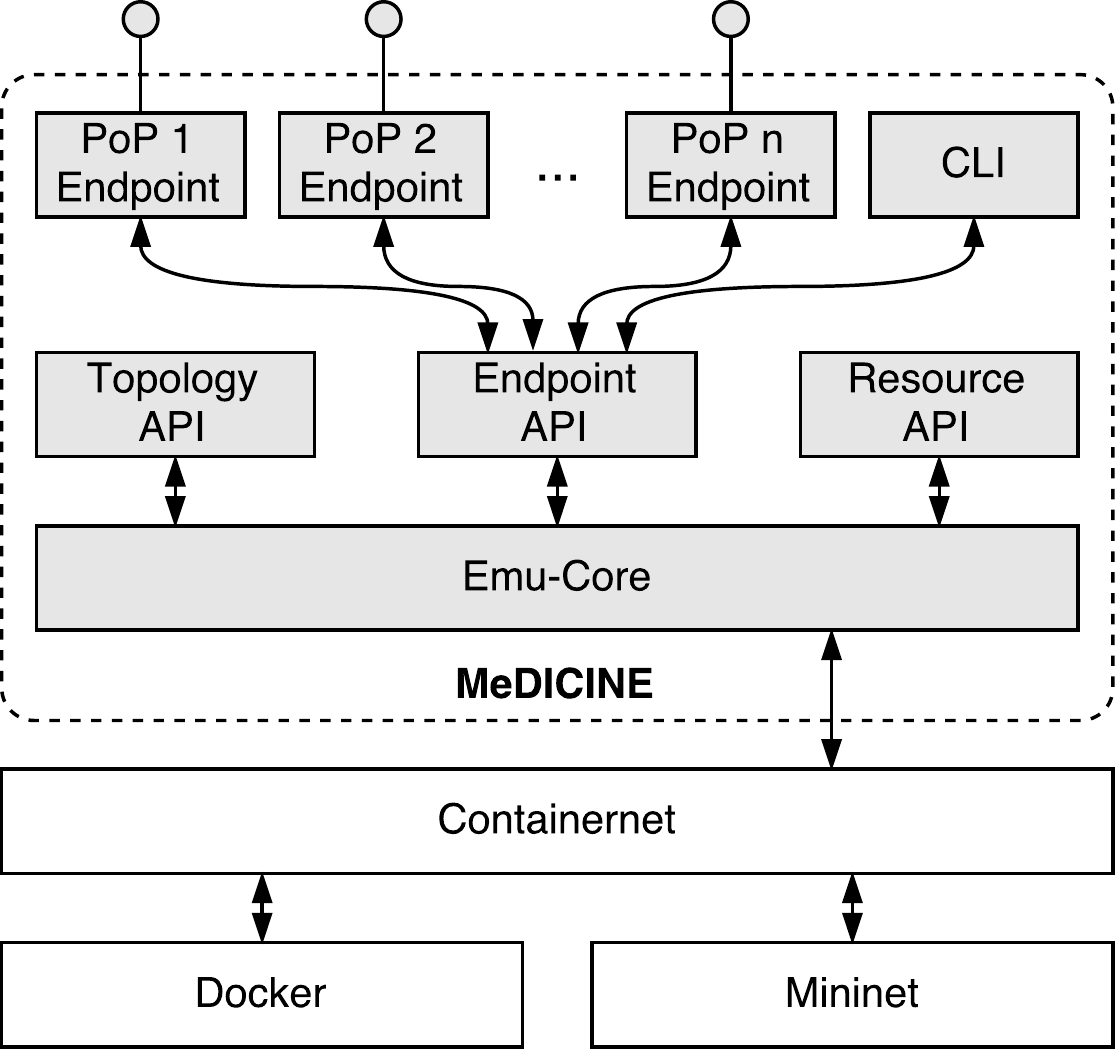}
  		\caption{System architecture and components with $N$ active PoP endpoints offering control interfaces to an external MANO system.}
  		\label{fig:medicine-components}
	\end{subfigure}
	\caption{The MeDICINE system}
\end{figure*}

%
%

\subsection{Background: Containernet}

We base the design of \emph{MeDICINE} on a tool called Containernet~\cite{Dockernet.webpage}. It extends the Mininet emulation framework and allows us to use standard Docker containers as compute instances within the emulated network. Containernet allows adding and removing containers from the emulated network at runtime, which is not possible in Mininet. This concept allows us to use Containernet like cloud infrastructure in which we can start and stop compute instances (in form of containers) at any point in time. Another feature of Containernet is that it allows to change resource limitations, e.g., CPU time available for a single container, at runtime and not only once when a container is started, like in normal Docker setups.

\subsection{Overview and Workflow}

To fulfill the previously defined use cases, \emph{MeDICINE} provides the following key features. First, it exploits Mininet's topology API to define arbitrarily complex multi-PoP environments with realistic link properties, like delay, bandwidth limitation, and loss rate. Second, it uses standard Docker containers to execute network functions, allowing a developer to directly deploy the prototyped container images to production PoPs after they have been tested locally. Third, it provides cloud-like interface endpoints, e.g., an OpenStack Heat-like interface, to control each emulated PoP in the platform. This allows developers to connect their local prototyping environment to existing MANO tools.

Fig.~\ref{fig:medicine-idea} shows the general idea of \emph{MeDICINE} and depicts the high-level workflow of a developer using it. First, the service developer defines a network service, consisting of function (VNFD) and service (NSD) descriptors as well as Docker files or pre-built images that contain the network functions to be tested~(1). The actual format of this network service and is descriptors depends on the used MANO system that deploys these services in our platform. Second, the developer defines a multi-PoP topology on which he wants to test the service~(2) and starts the \emph{MeDICINE} platform with this topology definition~(3). After the platform has been started, the developer connects the MANO system of his choice to the emulated PoPs by using a flexible endpoint API~(4) described in Sec.~\ref{sec:cloud-endpoints}. Now, the network service can be deployed on the platform by pushing it to the MANO system~(5) which starts each network function as a Docker container in an emulated PoP, connects it to the emulated network, and sets up its forwarding chain. Finally, the service is deployed and runs inside our platform~(6).

In this stage, a developer can directly interact with each running container through Containernets's interactive command line interface (CLI), e.g., to view log files, change configurations, or run arbitrary commands, while the service processes traffic generated by tools like \emph{iperf}. Furthermore, a developer and the MANO system can access arbitrary monitoring data generated by the platform or the network functions.

\subsection{System Architecture}
%
%
The system design of \emph{MeDICINE} is highly customizable. It offers plugin interfaces for most of its components, like API endpoints, container resource limitation models, or topology generators.

Fig. \ref{fig:medicine-components} shows the main components of our system and how they interact with each other. The \emph{emulator core} component implements the emulation environment, e.g., the emulated PoPs. It is the core of the system and interacts with the \emph{topology API} to load topology definitions. The flexible \emph{endpoint API} allows our system to be extended with different interfaces that can be used by MANO systems to manage and orchestrate emulated services (Sec. \ref{sec:cloud-endpoints}). The \emph{resource management API} allows to connect resource limitation models that define how much resources, like CPU time and memory, are available in each PoP (Sec.~\ref{sec:rm}). Finally, we provide an easy-to-use CLI that allows developers to interact with our platform.

\subsection{Topology Definition}
\label{sec:topology}

To test network services in realistic multi-PoP scenarios, test topologies define available PoPs, their resources, as well as the network and its properties between them. In contrast to classical Mininet topologies, our \emph{MeDICINE} topologies do not describe single network hosts connected to the emulated network but PoPs available in the network. 
In the most simplified case, such a PoP emulates just a single node, i.e., a router with attached compute and storage facilities like a Blade server. A more sophisticated node represents a small data center, which comprises several servers and is internally connected by a single SDN switch. More abstractly, an emulated PoP can also be a complex data center whose internal connection  is simplified into a “big-switch” abstraction (as shown in Fig. \ref{fig:medicine-idea}). For all these versions, we assume that the MANO system has full control over whether a particular VNF is executed at a particular PoP but does not care about internals of the PoPs.

A \emph{MeDICINE} topology allows to add an arbitrary number of SDN switches between PoPs (Fig.~\ref{fig:medicine-idea}, \emph{s1} and \emph{s2}). These SDN switches as well as any SDN switches within each PoP can be either controlled by standard SDN controllers, by custom controller implementations provided by the network service developer, or by the MANO system itself. Thus, complex network and forwarding setups with a high number of inter-PoP switches can be emulated.

We based our topology API on Mininet's Python API which has the benefit that developers can use scripts to define or algorithmically generate topologies. 
Listing \ref{lst:topo} shows an example topology script defining two PoPs that are interconnected by a single switch. It shows how the PoPs are connected and how the link setup is done (lines 1--8).

\subsection{Flexible Endpoint API}
\label{sec:cloud-endpoints}

After an emulation topology is defined, MANO systems need a way to start and stop compute instances within the emulated PoPs. To do so, we introduce the concept of \emph{flexible API endpoints} (Fig.~\ref{fig:medicine-components}). Such an API endpoint is an interface to a PoP that provides typical infrastructure-as-a-service (IaaS) semantics to manage compute instances. Instead of fixing our design to a single interface implementation, we provide an abstract API and allow users of the platform to implement their own endpoints on top of it. This has the benefit that our platform can be integrated with any MANO system as long as an API endpoint that provides the expected interfaces is created. Examples for such endpoints are OpenStack Nova or Heat-like interfaces, OpenVIM-like interfaces, or any other open or proprietary interface to which a MANO system should be connected.

These API endpoints are assigned to PoPs in the topology scripts (Listing~\ref{lst:topo} lines 15--20). The default approach is adding one endpoint to each PoP so that each emulated PoP provides its own management endpoint (Fig. \ref{fig:medicine-components}). From the perspective of the MANO system, this looks exactly like a real multi-PoP environment offering a heterogenous set of management interfaces towards the available PoPs.

\subsection{Networking and Chain Management}
\label{sec:chaining}

To emulate a fully working network service, we need to deploy its VNFs and set up the forwarding path between them, as shown in the service chain of Fig.~\ref{fig:medicine-idea}.
Since the emulated topology with PoPs consist of SDN switches, a service developer can have full control over the forwarding paths of the service's network traffic. Setting up a chain where traffic is steered along a defined path is now a matter of setting the correct forwarding entries in the SDN switches, with an SDN controller. 
To support developers, we provide a simplified API that brings Service Function Chaining (SFC) functionality into \emph{MeDICINE} while hiding the complexity of low-level SDN protocols. This API allows to chain running containers by calling a single method, i.e., \texttt{setChain(vnf1, ... , vnfN)}. Our solution uses VLAN tags as identifier to differentiate chains in multiple emulated services, similar to ongoing research regarding the use of Network Service Headers (NSH)~\cite{halpern2015service}.
An internal graph representation of the topology and its attached containers is kept to compute the forwarding chain with the fewest hops or the smallest delay.


%
%

%
%
\begin{lstlisting}[float=t, language=Python, label=lst:topo, caption=Example \emph{MeDICINE} topology with two PoPs connected to Heat-like cloud endpoints and example resource models.]
# create two PoPs
p1 = net.addPoP("My PoP 1")
p2 = net.addPoP("My PoP 2")
# create an intermediate SDN switch
s1 = net.addSwitch("s1")
# connect PoPs: p1 <-> s1 <-> p2
net.addLink(p1, s1, delay="10ms")
net.addLink(p2, s1, delay="50ms", loss=2)
# init. and assign resource models for each PoP
r1 = ResModelA(max_cu=24, max_mu=80, max_su=90)
r1.assignPoP(p1)
r2 = ResModelB(max_cu=80, max_mu=120, max_su=280)
r2.assignPoP(p2)
# instantiate and start cloud interface for each PoP
api1 = HeatCloudApiEndpoint(port=8004)
api1.connectPoP(p1)
api1.start()
api2 = HeatCloudApiEndpoint(port=8005)
api2.connectPoP(p2)
api2.start()
# run the emulation
net.start()
\end{lstlisting}

\subsection{Resource Models}
\label{sec:rm}

Even though cloud systems provide virtually infinite compute resources to their customers, realistic scenarios, especially with small PoPs, look different. Such PoPs offer limited compute, memory, and storage resources which have to be considered by a MANO system when placement and scaling decisions are taken. To emulate such resource limitations, \emph{MeDICINE} offers the concept of \emph{flexible resource models} assigned to each PoP (Listing~\ref{lst:topo} lines 10--13). These models are called whenever containers are allocated or released and they compute CPU time, memory, and storage limits for each of them. These models are also able to reject allocation requests, indicating that there are no free resources left on a given PoP. A generic API allows developers to easily create their own resource models. For example, a telco operator that deploys services in its own PoPs has more control about available resources than a web service provider that buys cloud resources from a third party, like Amazon. However, \emph{MeDICINE} can be used as a prototyping platform in both cases. The telco operator might, e.g., use a resource model in which resources are strictly reserved whereas the web service provider uses a model in which the service's performance is influenced by other random 3rd party services. Models for other operational metrics, like prizing models, can also be implemented, e.g., increase prizes for resources if a PoP is highly utilized.

To showcase how a \emph{MeDICINE} resource model looks like, we provide two example CPU limitation models that are implemented in our prototype; memory and storage models that use the same ideas are available as well. The goal of the presented models is to limit the overall available CPU capacity of each PoP in a way such that the utilization of one PoP does \emph{not influence} other PoPs. The presented models use the notion of \emph{compute units} (CU) to describe the relative amount of CPU time allocated to a single container. E.g., a container that requests $4$ CUs will get twice as much CPU time as a container requesting $2$ CUs. Using this, relative resource requirements can be described independently from absolute available resources. Further, we define:

\begin{itemize}
	\item$E_\textnormal{cpu} \in (0, 1]$ percentage of physical CPU time available for emulation. For example, all containers together will not use more than $60\%$ of the physical CPU if $E_\textnormal{cpu}=0.6$.
  	\item$N \in \mathbb{N}_{>0}$ number of PoPs in the emulated topology.
  	\item$\mathrm{mc}_p \in \mathbb{N}_{>0}$: number of \emph{CUs} available in PoP $p$.
	\item$\mathrm{ac}_p \in \mathbb{N}$: number of \emph{CUs} allocated in
  PoP $p$.
	\item$\mathrm{nc}_c \in \mathbb{N}_{>0}$: number of CUs requested for a container $c$.
	\item$P_c \in [0, 1]$ percentage of physical CPU time assigned to container $c$.
\end{itemize}

Based on this, we define a CPU limitation function as $f:~\mathrm{nc}_c~\times~p~\rightarrow~P_c$ and use it to introduce the following example resource models.

\subsubsection{Model A (Fixed Limit)}

Our first model assigns a fixed amount of available CUs to each PoP in the system. If not enough CUs are left in a PoP when a new container should be started, the instantiation request is rejected. As a result a PoP can never be over-utilized. Eq.~1 shows this model and how it computes the CPU time ($P_c$) assigned to a container that requests $\mathrm{nc}_c$ CUs. 

\begin{equation}
    f_p(\mathrm{nc}_c) =
	\begin{cases}
      \frac{E_{cpu}}{\sum_{i=1}^{N}{\mathrm{mc}_i}} \cdot \mathrm{nc}_c, & \text{if}\ \mathrm{ac}_p+\mathrm{nc}_c \leq \mathrm{mc}_p \\
      0\:\textnormal{\small (reject)}, & \text{else}
	\end{cases}
\end{equation}

\subsubsection{Model B (Cloud-like Over-Provisioning)}

Our second model does not have fixed CU limits per PoP. Instead, it allows CPU over-provisioning, which is a typical concept in IaaS clouds. For example, OpenStack Nova sets its default \emph{cpu\_allocation\_ratio} property to $16:1$ \cite{OpenStackNovaConf.wepage}. This results in situations in which allocated compute instances get less resources than initially requested. This means that a VNF might slow down if another VNF is started in the same PoP.

To model this behavior, we scale the limits of all containers within the same PoP by an over-provisioning factor defined as the fraction of available and currently used CUs in a PoP~(Eq.~2). Further, we update the limits of all containers of PoP $p$ whenever a new container is added or removed from $p$. As a result, the available CPU time for each container in an over-utilized PoP $p$ is reduced if more CUs than available are used (\(\mathrm{ac}_p > \mathrm{mc}_p\)). However, the limits of running containers in other PoPs are not changed and thus our model creates a realistic environment in which physically separated PoPs do not influence each other.

\begin{equation}
f_p(\mathrm{nc}_c) = \frac{E_{cpu}}{\sum_{i=1}^{N}{\mathrm{mc}_i}} \cdot \underbrace{\frac{\mathrm{mc}_p}{\max{\{\mathrm{mc}_p; \mathrm{ac}_p\}}}}_{\text{over-prov. factor}} \cdot \mathrm{nc}_c
\end{equation}

This provides a playground for realistic placement tests, e.g., an overloaded PoP might motivate a MANO system to reassign its containers to other PoPs with better performance. As a result, the previously over-utilized PoP is relieved and the performance of containers it is hosting improves.

Implementation-wise, our system does not use Docker's default CPU share limitation API since it is not sufficient for our use case. The first reason for this is that it only limits the CPU share if two or more containers want to utilize the entire CPU at the same time. It does not limit the CPU time if only one container is utilized and the competing ones are idle. Instead, we use the CPU bandwidth control functionalities of Linux's \emph{completely fair scheduler} (CFS) \cite{turner2010cpu}. The second reason for our custom implementation is that Docker's default API does only allow to set limitations when a container is started but not to update limits at runtime. We bypass these shortcomings by directly manipulating the \emph{cgroup} system properties. 

\section{Evaluation}
\label{sec:eval}

We did a couple of experiments to prove the overall concept, validate the behavior of the introduced resource models, and to showcase the system. The experiments use topologies with one and two PoPs in which Docker containers that run a workload generator (called \emph{stress}) are started. Thus, every container always tries to fully utilize the CPU. The overall available CPU time for containers is set to $E_\textnormal{cpu} = 0.5$ and the experiments are executed on a machine with Intel(R) Core(TM) i5-4690 CPU @ 3.50GHz and 16GB memory.

\begin{figure}[!ht]
\centering
\begin{subfigure}{.5\columnwidth}
  \centering
  \includegraphics[width=1.0\linewidth]{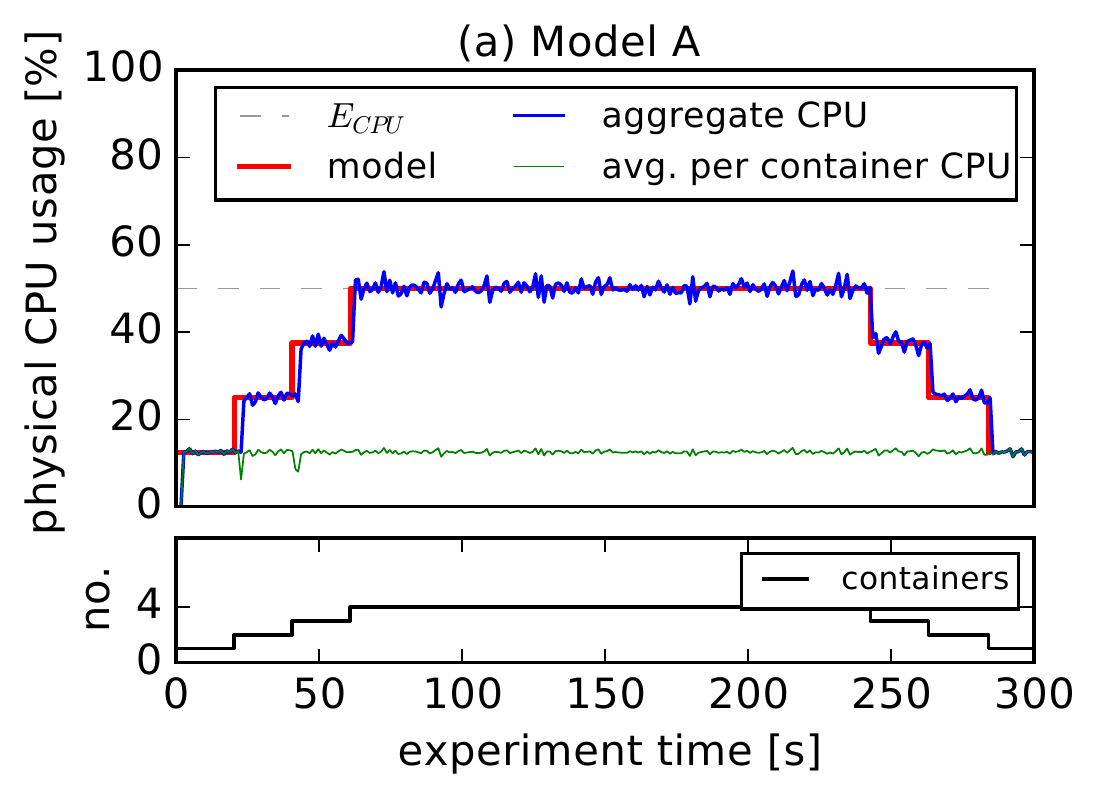}
  \vspace{-0.5cm}
  \label{fig:plot11}
\end{subfigure}%
\begin{subfigure}{.5\columnwidth}
  \centering
  \includegraphics[width=1.0\linewidth]{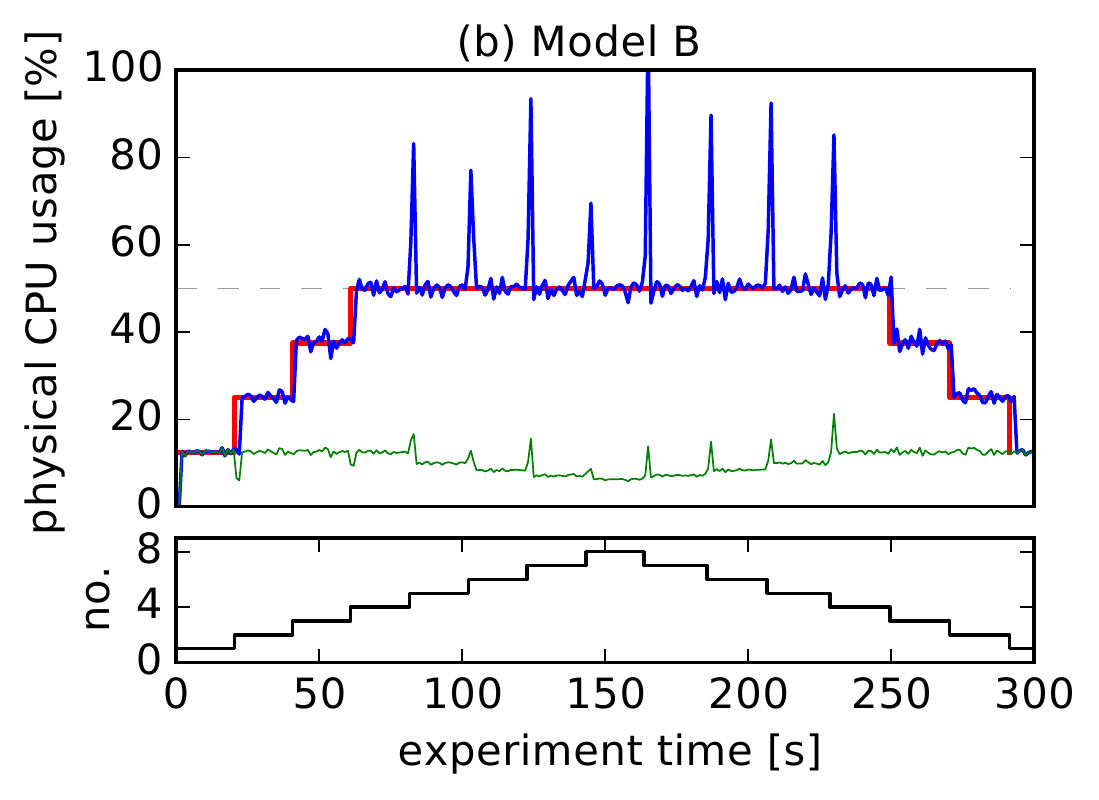}
  \vspace{-0.5cm}
  \label{fig:plot12}
\end{subfigure}
\caption{Modeled vs. measured container CPU usage}
\label{fig:plot1}
\end{figure}

The goal of our first experiment is to validate that the measured CPU usage of containers in a single PoP is aligned to the theoretical CPU usage computed by our models. During the experiment a new container (requesting $1$\,CU) is allocated every $20$\,s during the experiment until $8$ containers are requested. After additional $20$\,s, these containers are terminated one by one. The maximum limit of available CUs in the PoP is set to $4$\,CUs so that some of the requests are rejected (model A) or the PoP is over-utilized (model B). Fig.~\ref{fig:plot1} shows the aggregated CPU usage for the entire PoP as well as the average CPU usage for a single container, both measured with Docker's status API that returns detailed CPU time statistics. Additionally, the expected CPU utilization calculated by our models is plotted. The results show that the measured CPU utilization for all containers in the PoP is close to the limits computed by the model. The bottom graph in Fig.~\ref{fig:plot1}a shows how model A rejects requests after $4$ containers are allocated. Model B, in contrast, accepts all $8$ containers since it allows over-provisioning (Fig.~\ref{fig:plot1}b). It also shows how the available avg. CPU time for containers is reduced when the PoP is over-utilized. An interesting observation in Fig.~\ref{fig:plot1}b are the spikes in the measured CPU usage. We found that they happen during the reconfiguration of CPU limits of already running containers which is needed by model B and happens whenever a container is added or removed from the system. Investigating this issue and quantifying its effects will be part of future work.

\begin{figure}[!ht]
\centering
\begin{subfigure}{.265\columnwidth}
  \centering
  \includegraphics[width=1.0\linewidth]{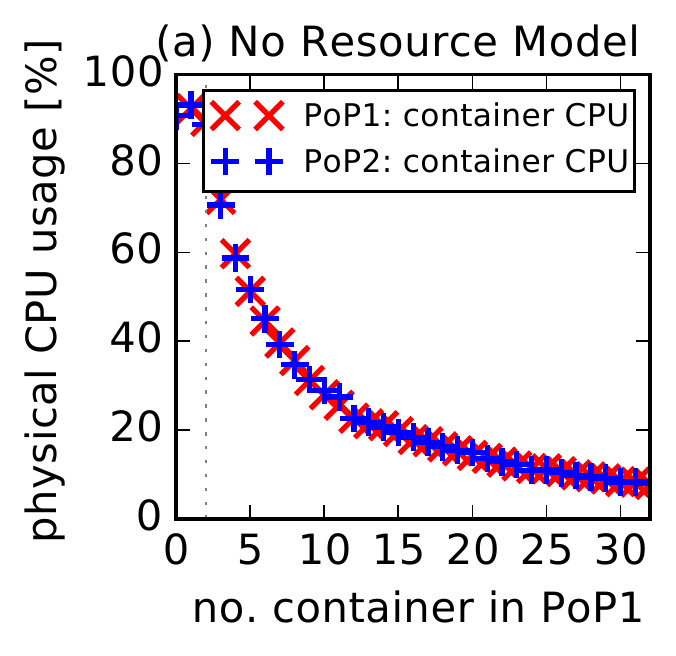}
  \vspace{-0.5cm}
  \label{fig:plot21}
\end{subfigure}%
\begin{subfigure}{.26\columnwidth}
  \centering
  \includegraphics[width=1.0\linewidth]{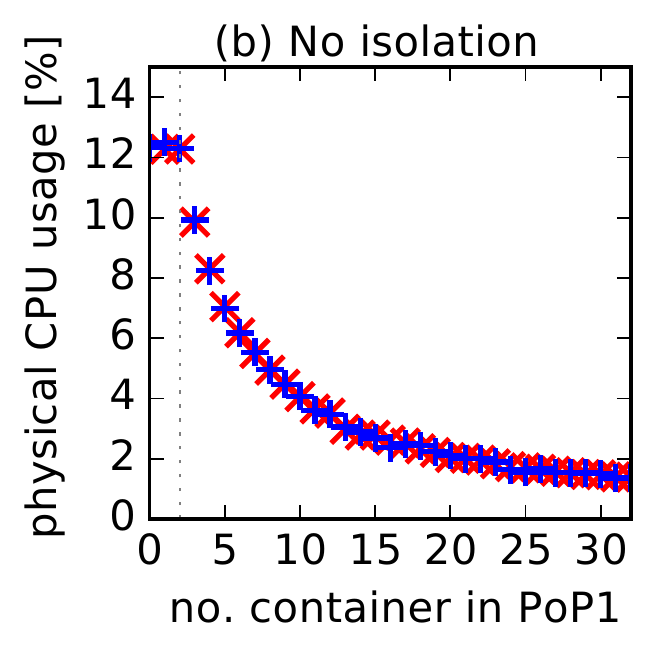}
  \vspace{-0.5cm}
  \label{fig:plot22}
\end{subfigure}%
\begin{subfigure}{.20\columnwidth}
  \centering
  \includegraphics[width=1.0\linewidth]{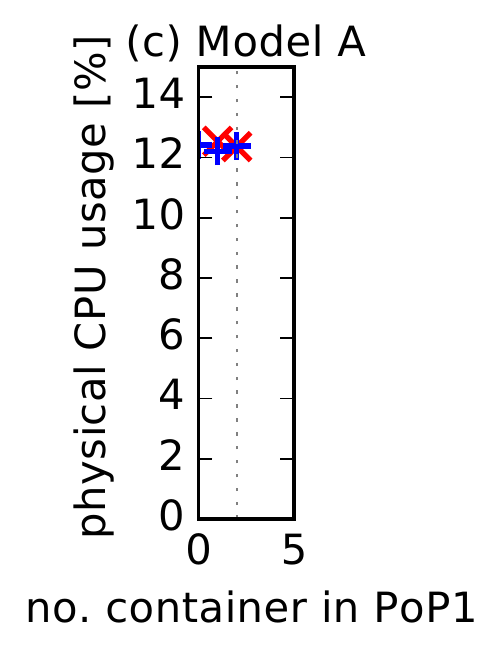}
  \vspace{-0.5cm}
  \label{fig:plot23}
\end{subfigure}
\begin{subfigure}{.26\columnwidth}
  \centering
  \includegraphics[width=1.0\linewidth]{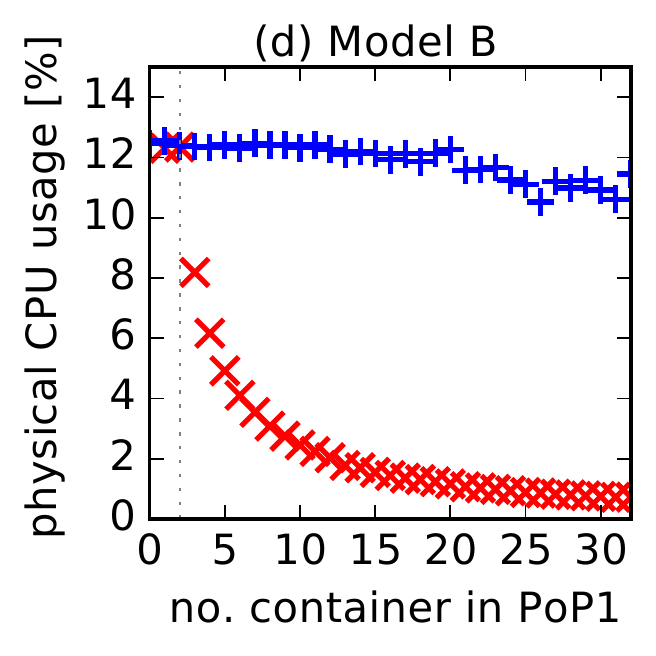}
  \vspace{-0.5cm}
  \label{fig:plot24}
\end{subfigure}
\caption{Cross-PoP resource isolation of resource models}
\label{fig:plot2}
\end{figure}

The second experiment shows how the presented resource models provide resource isolation between PoPs. It emulates a topology with two PoPs, each with a limit of $2$\,CUs. During the experiment, the avg. physical CPU time available for a single container is measured for different numbers of \emph {stress} containers running in PoP1. The number of \emph{stress} containers running in PoP2 is fixed to two running containers. With this, we can observe how the changing number of containers in PoP1 influences the performance of containers in PoP2. Fig.~\ref{fig:plot2}  shows the results for different resource models. Fig.~\ref{fig:plot2}a shows what happens when no resource limitation model is used. All containers compete for the entire physical CPU time and the performance of containers in PoP2 is reduced when more containers are added to PoP1. The same happens in Fig.~\ref{fig:plot2}b with the difference that it uses a common resource model for both PoPs. Thus, the containers do not influence each other until the maximum of two containers are running in PoP1. Fig.~\ref{fig:plot2}c shows the behavior of model A which does not allow over-provisioning and rejects all requests when two containers are already running in a PoP. The behavior of model B is shown in Fig.~\ref{fig:plot2}d. The figure validates that the model enables resource isolation between PoPs even when PoP1 is over-utilized and the CPU time for each of its containers is reduced. Fig.~\ref{fig:plot2}d shows that there are still some minimal influences to the performance of PoP2 when containers are added to PoP1, i.e., PoP2's performance is reduced by around 2\% when PoP1 is over-utilized by a factor of $16\times$.

\section{Conclusion and Future Work}
\label{sec:conclusion}

We introduced \emph{MeDICINE}, a novel prototyping platform for NFV that goes beyond its initial NFV use cases and is an excellent prototyping and test platform for distributed cloud services. 
The results of our experiments show that \emph{MeDICINE} can simulate resource limitations in multi-PoP environments while ensuring resource isolation between PoPs. 
By using standard Docker containers to execute network functions within our emulation platform, \emph{MeDICINE} allows developers to directly move their tested services into production environments without additional changes.

We believe that \emph{MeDICINE} is an important step towards a fully integrated development support toolchain for network service development. Its code is open-source and available as part of the emulation platform of the 5G-PPP project SONATA~\cite{sonemu.webpage}. We will continue our work in several directions, e.g., improved resource models with performance prediction, and advanced service chaining techniques.

The final version of this paper was accepted by~\cite{medicine2016}.

\section*{Acknowledgments}
\footnotesize{This work has been partially supported by the SONATA project, funded by the European Commission under Grant number 671517 through the Horizon 2020 and 5G-PPP programs (\url{www.sonata-nfv.eu}) and the German Research Foundation (DFG) within the Collaborative Research Centre ``On-The-Fly Computing" (SFB 901).}

\bibliographystyle{IEEEtran}
\bibliography{IEEEabrv,main}

\end{document}